\documentclass[lettersize,journal]{IEEEtran}
\usepackage{amsmath,amsfonts}
\usepackage{algorithmic}
\usepackage{algorithm}
\usepackage{array}
\usepackage[caption=false,font=normalsize,labelfont=sf,textfont=sf]{subfig}
\usepackage{textcomp}
\usepackage{stfloats}
\usepackage{url}
\usepackage{verbatim}
\usepackage{graphicx}
\usepackage{cite}
\usepackage{multirow}
\begin{document}

\title{An Interleaved Parallel Dependent Quantization Hardware Architecture for H.266/VVC}

\author{Jun Zhang, Weizhi Bian, Hao Zhang\textsuperscript{*}
	\thanks{Jun Zhang is with School of Automation, Central South University, Changsha 410083, China(email: junzhang@csu.edu.cn).}
	\thanks{Weizhi Bian is with School of Automation, Central South University, Changsha 410083, China(email: 234611099@csu.edu.cn).}
	\thanks{Hao Zhang is with School of Electronic Information, Central South University, Changsha, 410083, China(email: hao@csu.edu.cn)}
}

\markboth{An Interleaved Parallel Dependent Quantization Hardware Architecture for H.266/VVC}%
{Shell \MakeLowercase{\textit{et al.}}: A Sample Article Using IEEEtran.cls for IEEE Journals}

\IEEEpubid{}

\maketitle

\begin{abstract}
While dependent quantization in H.266/VVC delivers a high compression ratio, its strong serial nature and high complexity result in poor real-time performance, making it difficult to deploy in practical scenarios. To improve the real-time performance of dependent quantization with minimal degradation to its compression performance, we propose a interleaved parallel dependent quantization hardware architecture with low BDBR loss, which achieves four-channel parallel dependent quantization by time-division multiplexing most combinational logic. This architecture adopts the proposed intra-CG context simplification scheme and the encoding scheme that skips decAbslevel during rate estimation. The proposed design incurs a BDBR loss of only 0.42\% under the All Intra configuration and 0.38\% under the Random Access configuration, respectively. Implemented in Verilog HDL, the dependent quantization hardware architecture achieves quantization speeds of 4K@33.2, 91.4, 331.2, and 454.1 fps at QP = 22, 27, 32, and 37, respectively, when implemented on the Xilinx XCZU19 FPGA. When implemented on ASIC using the TSMC 28nm process standard cell library, the corresponding quantization speeds reach 4K@84.0, 231.0, 837.0, and 1147.7 fps.
\end{abstract}


\section{Introduction}

\IEEEPARstart{N}{owadays}, the efficient and intelligent video viewing experience benefits not only from the increased data transmission bandwidth, but also from the continuous development of video codec technologies. With the growing popularity of ultra-high-definition videos, 360° panoramic videos, AR/VR videos and remote conferences, video coding standards such as H.264/AVC and H.265/HEVC have become difficult to meet the current video compression requirements. For this reason, the H.266/VVC video coding standard was officially finalized in July 2020\cite{R:B. Bross}.

Compared with its predecessor standard H.265/HEVC, H.266/VVC incorporates a host of novel coding tools to enhance video compression performance, including more flexible block partitioning schemes\cite{R:Block partitioning}, richer intra/inter prediction modes\cite{R:Intra prediction, R:Inter prediction}, a wider array of transform and quantization tools\cite{R:Transform, R:Quantization}, and more adaptive loop filtering\cite{R:Loop Filter}. While these newly introduced coding tools enable H.266/VVC to achieve a 50\% higher compression ratio compared to H.265/HEVC, they also lead to a substantial increase in encoding complexity. This not only renders real-time encoding of H.266/VVC infeasible, but also impedes its large-scale deployment in practical application scenarios. Among these factors, quantization is a critical reason that hinders the encoder from achieving real-time encoding performance.

The quantization tool is the fundamental cause of lossiness in audio and video compression. It plays a critical role in audio and video compression and directly affects the coding quality and encoding complexity of audio and video content. The quantization methods in the H.26x series mainly include Uniform Scalar Quantization (USQ), Rate-Distortion Optimized Quantization (RDOQ) and Dependent Quantization (DQ). Among them, DQ is a newly introduced quantization method in the H.266/VVC standard. Compared with its predecessor quantization method RDOQ, it achieves approximately a 2\% BD-Rate reduction, while introducing a high degree of seriality. 

To address the issue of high seriality in the quantization process that impedes real-time quantization and improve quantization speed, numerous researchers have conducted studies on RDOQ and DQ.

For the research on RDOQ, Lee et al. \cite{R:H. Lee} proposed a method for RDOQ bypass decision-making and quantization candidate simplification via statistical analysis, which reduces the average quantization complexity by 14.8\% with almost no loss in coding performance. Canh et al. \cite{R:T. Nguyen Canh} pioneered the application of deep neural networks to RDOQ, taking the results of scalar quantization as input to directly predict the optimal quantization level without calculating bitrate and distortion. The experimental results demonstrate that this scheme has promising potential performance. Cui et al. \cite{R:J. Cui} adopted a mixed Laplace distribution to model bitrate and distortion, thereby simplifying their calculations. This approach cuts down quantization time by 70\% while only causing a 0.3\%–0.4\% drop in coding efficiency. Xu et al. \cite{R:M. Xu.RDOQ} proposed a quantization level estimation scheme to simplify RDOQ computation, which saves quantization time by 14.39\% (AI-main), 12.88\% (RA-main) and 12.91\% (LB-main) with nearly no Bjøntegaard Delta Bit Rate (BDBR) loss.

For the research on DQ, Xu et al. \cite{R:M. Xu.DQ} adopted a binary classifier to determine the DQ starting point, and analyzed the relationship between pre-quantization values and the state of the previous coefficient to prune partial trellis branches, thereby reducing the software encoding complexity. Wang et al. \cite{R:M. Wang} reduced the DQ complexity by adaptively adjusting the trellis starting point and pruning unnecessary trellis paths.

The aforementioned research on quantization has focused primarily on software acceleration. As for hardware-oriented research on quantization, it mainly includes the following aspects: Sheng et al. \cite{R:Q. SHENG} proposed a hardware implementation method for DQ based on the parallel Viterbi algorithm, which achieves a processing speed of 1080P@60Hz with a BDBR loss of 1.05\% (LB) and 1.03\% (RA). Zhang et al. \cite{R:J. Zhang} proposed a simplified rate-distortion calculation scheme and implemented a coefficient-level pipelined DQ hardware architecture, which achieves a quantization speed of approximately 4K@30fps with a BDBR loss of around 0.8\%. Although these two studies have improved the encoding speed of dependent quantization, they have also resulted in a significant degradation in BDBR performance. Therefore, it is highly necessary to research a hardware architecture with high parallelism and low BD-Rate loss for H.266/VVC dependent quantization.

The remaining part of this paper is organized as follows. Chapter II introduces the basic principles of dependent quantization involved in this paper, including the fundamental principles of dependent quantization and the principles of rate-distortion cost calculation. Chapter III presents the proposed improvements to DQ. Chapter IV provides a detailed description of the proposed hardware parallel architecture. Chapter V presents the experimental results and comparisons. The final conclusion of this paper is summarized in Chapter VI.

\section{Dependent Quantization in H.266/VVC}
In this chapter, we first present an overview of the overall situation of dependent quantization, and then discuss the challenges encountered in the implementation of dependent quantization hardware architecture.

\subsection{Overview of Dependent Quantization In H.266/VVC}
The concept of dependent quantization originated from trellis coding quantization (TCQ)\cite{R:M. W.} and is a new quantization tool introduced in the video codec standard. Its essence is a high-dimensional vector quantization.

\begin{figure*}[!t]
	\centering
	\subfloat[]{\includegraphics[width=5in]{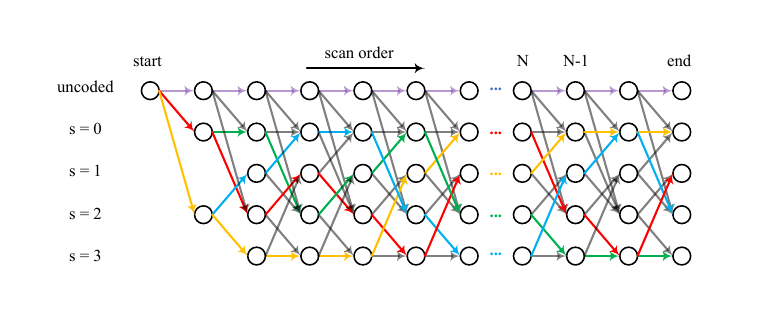}}
	\hfill
	\subfloat[]{\includegraphics[width=2in]{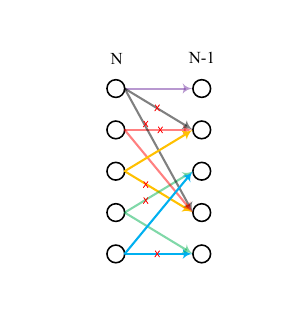}}
	\caption{The trellis structure of the encoder side.}
	\label{F.Trellis_structure}
\end{figure*}

Dependent Quantization operates at the transform block (TB) level. The quantization process is illustrated in Fig. \ref{F.Trellis_structure}(a). Essentially, the quantization procedure can be interpreted as the construction of a trellis structure, where each column of nodes represents multiple quantization states corresponding to a single transform coefficient. The number of columns in the trellis corresponds to the number of transform coefficients within a TB that require quantization. 

For the trellis structure constructed during the quantization of a TB, as illustrated in Fig. \ref{F.Trellis_structure}(a), only a single optimal quantization path is ultimately selected for output. This selection is carried out during the quantization process using the Viterbi algorithm.

In Fig. \ref{F.Trellis_structure}(a), the five colored lines represent the five paths with the smallest accumulated rate–distortion costs that are retained during the quantization process. As shown in Fig. \ref{F.Trellis_structure}(b), the pruning procedure of the Viterbi algorithm between the N-th and (N-1)-th columns is illustrated. For each incoming branch corresponding to the five states of the (N-1)-th coefficient, the accumulated RD costs are compared, and the branches with larger RD costs are pruned. The pruned paths are marked with “×” in the figure, while only the path with the minimum RD cost is preserved.

By iteratively performing the procedure shown in Fig. \ref{F.Trellis_structure}(b), five candidate quantization paths are eventually obtained for each TB, as depicted in Fig. \ref{F.Trellis_structure}(a). Among these five paths, the one with the minimum rate–distortion cost is selected as the final quantization result.

Each transform coefficient in the dependent quantization process corresponds to 4 quantization candidates. Prior to calculating these quantization candidates, a pre-quantization process needs to be performed on the original transform coefficients, which is expressed as Equation \ref{E.preQuant}. 

\begin{equation}
	\begin{aligned}
		qIdx=\max\left(1,\lfloor\frac{|TCoeff|}{\Delta}+\theta\rfloor\right)
	\end{aligned}
	\label{E.preQuant}
\end{equation}

$qIdx$ denotes the pre-quantization result, $TCoeff$ represents the transform coefficient to be quantized, $\Delta$ denotes the quantization step size that is related to the quantization parameter $QP$, and $\theta$ represents the offset.

The calculation method for the quantization candidates is expressed as Equation \ref{E.CandidateCalc}. After the calculation using Equation \ref{E.CandidateCalc}, a total of four quantization candidates, namely $Qc_0$ to $Qc_3$, will be obtained.

\begin{equation}
	\begin{aligned}
		Qc_x=(qIdx + x + 1) >> 1, x = 0, 1, 2, 3
	\end{aligned}
	\label{E.CandidateCalc}
\end{equation}

During the optimal quantization path selection process illustrated in Fig. \ref{F.Trellis_structure}, the selection criterion is the accumulated rate–distortion cost along each path. For all quantization candidates of each transform coefficient, their rate–distortion costs must be computed in real time during the quantization process. The formulation of the rate–distortion cost is given in \ref{E.RdcostCalc}.

\begin{equation}
	J = D + \lambda\cdot R
	\label{E.RdcostCalc}
\end{equation}

$\lambda$ is the Lagrange multiplier, which depends on the quantization parameter QP. D represents distortion, and R represents rate.

Distortion $D$ is typically represented using Mean-Square Error (MSE), and the distortion calculation for quantization level value can be expressed as Equation \ref{E.DistortionCalc} \cite{R:H. Schwarz}.
\begin{equation}
	D(q_k,s_k) = (t_k-(2\cdot Q_c-(s_k\gg 1)\cdot sgn(Q_c))\cdot \triangle)^2.
	\label{E.DistortionCalc}
\end{equation}

$t_k$ represents the original transform coefficient, $Q_c$ represents the quantization level, $s_k$ represents the quantization state of dependent quantization, and $\triangle$ denotes the quantization step.

The rate $R$ is derived from the rate estimation process. Similar to CABAC encoding, the rate estimation process also requires constructing a context model for each quantization path before performing rate estimation on the individual syntax elements decomposed from the quantization candidates. The sum of the rate estimates of these syntax elements is taken as the estimated result of the quantization candidates, as expressed in Algorithm \ref{Alg.RateEstimate}.

\begin{algorithm}[htbp]
	\caption{Rate Estimate process.}
	\label{Alg.RateEstimate}
	\begin{algorithmic}
		\STATE
		\STATE {if $codedSyntax < Mccb - 4$} 
		\STATE {$\quad R_{Qc} = R_{sig} + R_{par} + R_{gt1} + R_{gt3} + R_{rem}.$}
		\STATE {else}
		\STATE {$\quad R_{Qc} = R_{decAbslevel}.$}
	\end{algorithmic}
\end{algorithm}

$sig$, $par$, $gt1$, $gt3$ and $rem$ denote syntax elements, $codedSyntax$ represents the number of coded syntax elements, the calculation of $Mccb$ is expressed as Equation \ref{E.MccbCalc}, the decomposition of 
$Qc$ into syntax elements is given by Equation \ref{E.QcRateCalc}, and $decAbslevel$ indicates the amplitude of $Qc$
with its calculation process shown in Equation \ref{E.decAbslevelCalc}.

\begin{equation}
	Mccb = (TbWidth \times TbHeight) \times 1.75.
	\label{E.MccbCalc}
\end{equation}

\begin{equation}
	Qc = sig + par + gt1 + 2 \cdot gt3 + 2\cdot rem.
	\label{E.QcRateCalc}
\end{equation}

\begin{equation}
	\begin{aligned}
		decAbslevel &= \left\{
		\begin{array}{ll}
			ZeroPos,&Qc = 0 \\
			Qc-1,&Qc \leq ZeroPos \\
			Qc,&Qc > ZeroPos
		\end{array}
		\right.	\\
		ZeroPos &= 2^m\cdot\left\{
		\begin{array}{ll}
			1,& Qstate < 2 \\
			2,& Qstate \geq 2
		\end{array}
		\right.	
	\end{aligned}
	\label{E.decAbslevelCalc}
\end{equation}

$m$ represents the Rice parameter for Golomb-Rice coding, which can be calculated in Equation \ref{E.RiceParCalc}.
\begin{equation}
	\begin{aligned}
		m &= \left\{
		\begin{array}{ll}
			0,& s_T < 7 \\
			1,& 7 \leq s_T < 14 \\
			2,& 14 \leq s_T < 28 \\
			3,& s_T \geq 28
		\end{array}
		\right.	\\
		s_T &= \sum_{T} Qc - 5\cdot z_0
	\end{aligned}
	\label{E.RiceParCalc}
\end{equation}

During the computation of the Rice parameter for $rem$, the value assigned to $z_0$ is 4. When determining the Rice parameter for $decAbslevel$, $z_0$ is set to 0. When directly using bypass mode to encode quantization levels, the rate of the quantization level is calculated as.

\begin{figure}[!t]
	\centering
	\includegraphics[width=2.5in]{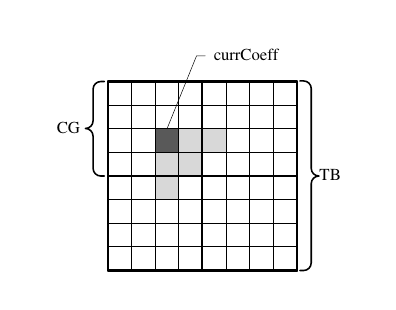}
	\caption{context template}
	\label{F.context_template}
\end{figure}

Among these, the rate estimation processes for $sig$, $par$, $gt1$, $gt3$ and the Rice parameter 
$m$ all rely on the context template illustrated in Fig. \ref{F.context_template}. That is to say, estimating the rate of the syntax elements of the current coefficient (in solid black) requires prior knowledge of the quantization results of other coefficients within the context template.

\subsection{Hardware Design Challenge Analysis}
\begin{figure}[!t]
	\centering
	\includegraphics[width=3.5in]{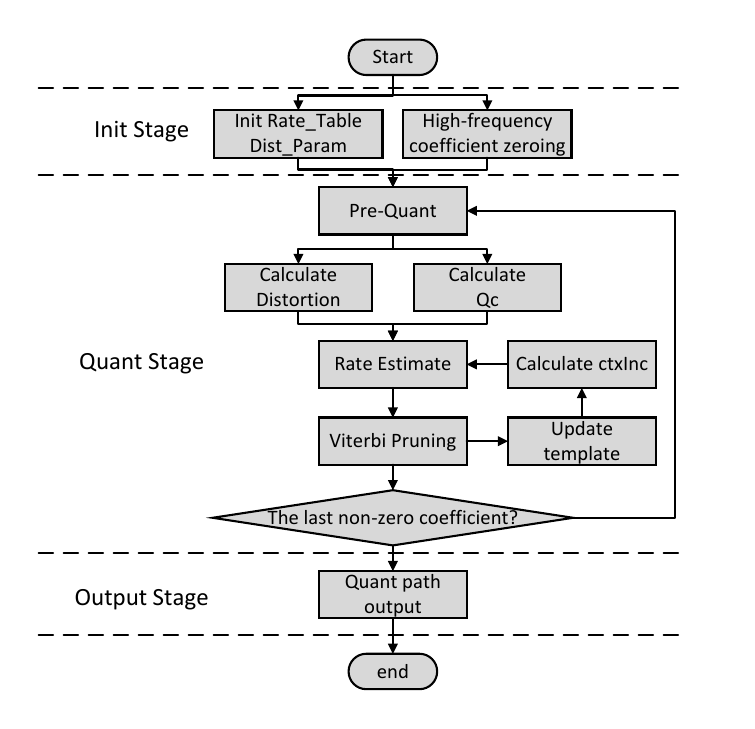}
	\caption{Dependent Quantization Flowchart}
	\label{F.DQFlowChart}
\end{figure}

Based on the analysis presented earlier, the flowchart of the dependent quantization process is illustrated in Fig. \ref{F.DQFlowChart}. The dependent quantization process is mainly divided into three stages, namely the initialization stage, the quantization stage and the output stage. Among these stages, the quantization stage is the primary factor that impedes the hardware parallelization of dependent quantization.

As can be seen from Fig. \ref{F.DQFlowChart}, the rate estimation process of the current coefficient relies on ctxInc calculation, which is related to the quantization results of previous coefficients. In other words, the quantization of the current coefficient cannot proceed until the quantization of the previous coefficient is completed. The quantization stage consists of 7 steps in total. Assuming that each step can be completed within one clock cycle and that Calculate Distortion and Calculate QIdx can be executed in parallel, quantifying a single transform coefficient will take 6 clock cycles. Excluding the Init Stage and Output Stage, quantifying a single 4K (3840×2160) frame requires $(6 \times 3840 \times 2160)$ clock cycles. Given a hardware clock frequency of 200 MHz, the maximum achievable frame rate of the hardware is only $(200M \div (6 \times 3840 \times 2160)=4fps)$, which is far from meeting the performance requirements for real-time quantization of 4K video.

The above analysis assumes that each step in the Quant Stage (as shown in Fig. \ref{F.DQFlowChart}) can be completed within one clock cycle. However, in the actual dependent quantization process, the Calculate ctxInc step is difficult to finish in a single clock cycle. This is because this process requires multiple data fetches from the context template illustrated in Fig. \ref{F.context_template}, involving up to 5 memory access operations, which will significantly increase the time required to quantize a single coefficient in the Quant Stage.

In addition to the above challenges regarding hardware quantization speed, there are also formidable challenges in terms of hardware resource consumption. As illustrated in Fig. \ref{F.Trellis_structure}, the quantization process of each TB block involves 4 quantization paths, which means that 4 sets of context templates (as shown in Fig. \ref{F.context_template}) need to be maintained. This will introduce a memory resource overhead of 4 blocks of 
$32 \times 32$ byte (since each TB block contains a maximum of $32 \times 32$ coefficients).

Based on the above analysis, due to the highly serial nature inherent in dependent quantization, the hardware architecture design of dependent quantization is confronted with the following key challenges:
(1) The transform coefficients can only be quantized sequentially one after another, making parallelization difficult to achieve.
(2) The rate estimation process for the current transform coefficient cannot be initiated until the quantization of the preceding transform coefficients is completed, which results in a large number of waiting cycles during the hardware quantization process.
(3) The four quantization paths constructed in the trellis diagram for the quantization process will lead to substantial memory resource consumption and a large number of memory access operations, which not only poses a challenge to hardware resources but also increases the latency of each transform coefficient in the Quant Stage.

\section{Proposed Parallelized Dependent Quantization Scheme}
To address the challenges faced by the dependent quantization hardware design mentioned in Chapter II, we propose the following two improvements to reduce the complexity of the dependent quantization hardware architecture, thereby balancing the performance and resource consumption of the dependent quantization hardware architecture. The two schemes are as follows: 

\subsection{Intra-CG context template}
To address the problems mentioned in Chapter 2—where the rate estimation process incurs multi-cycle memory accesses that degrade quantization speed, and the maintenance of multiple quantization paths leads to substantial resource consumption—this paper proposes a scheme that confines the context templates required for ctxInc calculation within a single CG (Coefficient Group). During the rate estimation of dependent quantization, it is necessary to calculate the ctxInc of each syntax element described in Chapter 2. The computation of these context indices requires accessing multiple already quantized CGs within the same TB (Transform Block), as illustrated in Fig. \ref{F.context_template}. Cross-CG access gives rise to two key issues:
(1) The quantization process must use BRAM (Block Random Access Memory) to store the quantization results of all coefficients in the TB, resulting in high resource consumption.
(2) Due to the use of BRAM for storage, each ctxInc calculation requires multiple memory access operations, which significantly increases the time required to quantize a single coefficient.

\begin{figure}[!t]
	\centering
	\includegraphics[width=3.5in]{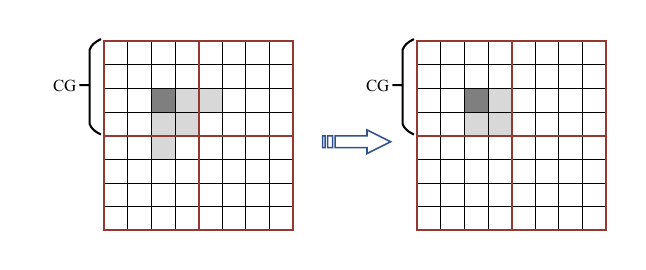}
	\caption{Intra-CG context template}
	\label{F.Intra-CG}
\end{figure}

The intra-CG context scheme proposed in this paper can address the two aforementioned issues. The ctxInc of each coefficient is only related to the already quantized coefficients within the CG where the current coefficient resides, and the positions in the context template that fall outside this CG are filled with 0, as illustrated in Fig. \ref{F.Intra-CG}.

The intra-CG context scheme offers the following two advantages:
(1) Only the data of a single CG needs to be stored, and merely 16 registers are sufficient to maintain the ctxInc of one quantization path.
(2) Since registers are adopted for storage, data from multiple context templates can be read simultaneously within one clock cycle, eliminating the need for multi-cycle memory access operations and reducing the quantization latency of a single coefficient.

To verify the feasibility of the intra-CG context scheme, we conducted ablation experiments on it. Three video sequences with different resolutions from the CTC (Common Test Conditions) were selected for the experiments, namely BQSquare, RaceHorsesC, and Cactus. The BDBR (Bjøntegaard Delta Bit Rate) of the three sequences using the intra-CG context scheme was tested under both All Intra and Random Access configurations, with the experimental results presented in Table I.
The results in Table \ref{T.Intra_CG_ablationExp} demonstrate that while the proposed scheme drastically reduces the number of BRAMs and the quantization latency of a single coefficient, it only incurs a BDBR loss of 0.3\%. This paper argues that such a loss falls within an acceptable range, and thus the subsequent hardware architecture design adopts this scheme.

\begin{table}[!t]
	\caption{BDBR(\%) OF THE INTRA‑CG CONTEXT SCHEME AGAINST\\ VTM 18.0.}
	\label{T.Intra_CG_ablationExp}
	\centering
	\setlength{\tabcolsep}{4mm}{
		\begin{tabular}{c|c|c|c}
			\hline
			Sequence & Resolution & All Intra & Random Access \\
			\hline
			Cactus & 1920x1080 & 0.39 & 0.44 \\
			RaceHorsesC & 832x480 & 0.47 & 0.57 \\
			BQSquare & 416x240 & 0.32 & 0.30 \\
			\hline
			\multicolumn{2}{c|}{Average} & 0.39 & 0.44 \\
			\hline
	\end{tabular}}
\end{table}

\subsection{Skipping the syntax element $decAbslevel$}
As can be seen in Algorithm \ref{Alg.RateEstimate}, in CABAC encoding, when the number of coded syntax elements of a TB block exceeds $Mccb-4$, CABAC encoding will adopt the bypass coding mode for the coefficient amplitude $decAbslevel$. The purpose of this design in the H.266/VVC video coding and decoding standard is to reduce the software complexity of CABAC encoding. This is because the bypass coding of $decAbslevel$ does not require real-time context updating and is relatively simpler. The video coding and decoding reference software VTM18.0 adopts the same method to perform rate estimation for the transform coefficients in dependent quantization.

However, in the design of the dependent quantization hardware architecture, a unified rate estimation method for transform coefficients not only does not increase hardware complexity, but also reduces some logic and saves hardware resources. Based on the above considerations, this paper conducts ablation experiments on the scheme of skipping $decAbslevel$ coding to further analyze whether it is necessary to retain $decAbslevel$ coding. The experimental conditions are consistent with those in Chapter-III-A, and the experimental results are presented in Table \ref{T.decAbslevel_ablationExp}.

\begin{table}[!t]
	\caption{BDBR(\%) OF THE SKIPPING DECABSLEVEL SCHEME AGAINST VTM18.0.}
	\label{T.decAbslevel_ablationExp}
	\centering
	\setlength{\tabcolsep}{4mm}{
		\begin{tabular}{c|c|c|c}
			\hline
			Sequence & Resolution & All Intra & Random Access \\
			\hline
			Cactus & 1920x1080 & 0.00 & 0.00 \\
			RaceHorsesC & 832x480 & 0.03 & 0.01 \\
			BQSquare & 416x240 & 0.09 & 0.15 \\
			\hline
			\multicolumn{2}{c|}{Average} & 0.04 & 0.05 \\
			\hline
	\end{tabular}}
\end{table}

The results in Table \ref{T.decAbslevel_ablationExp} demonstrate that skipping the coding of $decAbslevel$ only incurs a BDBR loss of 0.1\%, while it can reduce the consumption of certain hardware resources. Therefore, this paper adopts this scheme for the subsequent design of the dependent quantization hardware architecture.

\section{Interleaved Parallel Dependent Quantization Hardware Architecture Design}

\subsection{Zeroing high-frequency coefficients}
A transform block (TB) that has undergone discrete cosine transform (DCT) separates the high-frequency and low-frequency coefficients of an image, with low-frequency coefficients located at the top-left corner of the TB and high-frequency coefficients located at the bottom-right corner. Since the human eye is insensitive to the high-frequency components of an image, zeroing out high-frequency coefficients barely affects the visual quality of the image. Therefore, a large number of high-frequency coefficients are zeroed during the quantization process. In the dependent quantization process, the high-frequency coefficient zeroing operation is essentially a process of searching for the first non-zero coefficient in reverse zig-zag order, traversing from the bottom-right corner to the top-left corner of the TB. This process is divided into two steps: the first step is to perform a rough search prior to the Quant Stage, which corresponds to the Zeroing high-Freq Coeff module in Fig. \ref{F.DQFlowChart}. The core of this search process is to sequentially compare transform coefficients with a preset threshold in the defined scanning order; the first transform coefficient that exceeds the threshold is identified as the first non-zero quantized coefficient.

\begin{figure}[!t]
	\centering
	\includegraphics[width=4in]{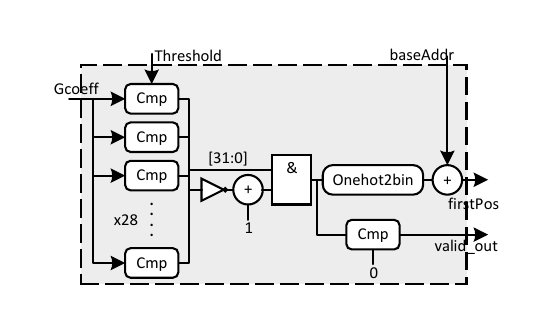}
	\caption{High Frequency Zeroing}
	\label{F.HighFreqZeroing}
\end{figure}

To improve the parallelism of the hardware module for high-frequency coefficient zeroing, this paper compares 32 coefficients with the Threshold simultaneously within one clock cycle. The process of high-frequency coefficient zeroing is presented in Algorithm \ref{Alg.HighFreqZeroing}, and its hardware architecture is illustrated in Fig. \ref{F.HighFreqZeroing}. This hardware architecture can ensure that the high-frequency coefficient zeroing operation for 32×32 transform coefficients is completed within 32 clock cycles. Given that the maximum size of a transform block (TB) is 64×64, and in dependent quantization, only the 32×32 low-frequency coefficients in the top-right corner of TBs with sizes of 64×64, 32×64, and 64×32 are quantized, this architecture is capable of completing the high-frequency coefficient zeroing operation for TBs of any size within 32 clock cycles.

\begin{algorithm}[htbp]
	\caption{Zeroing high-frequency coefficients.}
	\label{Alg.HighFreqZeroing}
	\begin{algorithmic}
		\STATE
		\STATE {\textbf{Step 1:} 32 coefficients (denoted as Gcoeff) are compared with the threshold value Threshold simultaneously. A comparison result of 1 is generated if a coefficient is greater than or equal to Threshold; otherwise, the result is 0.} 
		\STATE
		\STATE {\textbf{Step 2}: The 32-bit value isGreatZero obtained from Step 1 is used to calculate the one-hot code corresponding to the position of the first non-zero coefficient from right to left, in accordance with the formula below.}
		\STATE \begin{center}{\textbf{Pos\_onehot = isGreatZero \& ( $\sim$ isGreatZero + 1)}}\end{center}
		\STATE
		\STATE {\textbf{Step 3}: 
			\\Convert the one-hot code into a binary code: \\
			\begin{center}{\textbf{Pos\_onehot $\rightarrow$ Pos\_bin}}\end{center}
			The base address (baseAddr) of the current 32 coefficients plus the position (Pos\_bin) of the first non-zero coefficient within these 32 coefficients: \\
			\begin{center}{\textbf{firstNonZeroCoeff = Pos\_bin + baseAddr}}\end{center}}
		\STATE {As for \textbf{valid\_out}:}
		\STATE {\quad if $Pos\_onehot = 0$} 
		\STATE {\quad \quad $valid\_out = 0.$}
		\STATE {\quad else}
		\STATE {\quad \quad $valid\_out = 1.$}	
		\STATE {A value of valid\_out = 0 indicates that firstNonZeroCoeff is valid, and the high-frequency coefficient zeroing operation is terminated. Otherwise, firstNonZeroCoeff is invalid, and the search continues in the next GCoeff.}
	\end{algorithmic}
	\label{alg2}
\end{algorithm}

\subsection{4-channel Parallel Hardware Dependent Quantization Architecture}

Based on the dependent quantization optimization scheme proposed in Chapter III, the parallel hardware architecture for dependent quantization designed by us is illustrated in Fig. \ref{F.arch}. As illustrated in Fig. \ref{F.DQFlowChart}, the process of quantizing a Transform Block (TB) is mainly divided into three stages:

\begin{figure*}[!t]
	\centering
	\includegraphics[width=7in]{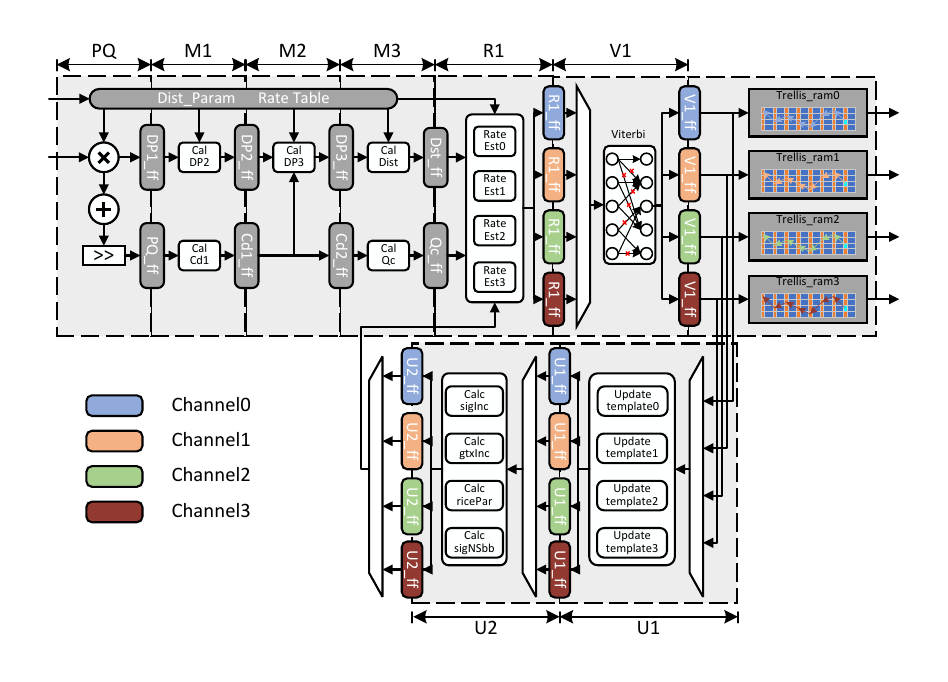}
	\caption{An Interleaved Parallel Dependent Quantization Hardware Architecture for H.266/VVC}
	\label{F.arch}
\end{figure*}

Init Stage: This stage completes the initialization of rate tables and distortion parameters, as well as the zeroing of high-frequency coefficients. It takes a fixed 32 clock cycles to finish.

Quant Stage: The initialized TB is quantized in this stage. Since the position of the first non-zero coefficient varies from one TB to another, the number of clock cycles required to quantize TBs of the same size differs.

Output Stage: From the four quantization paths in the trellis data constructed during the quantization stage, the trellis path with the minimum rate-distortion cost is selected, and the quantization results are output in sequence according to the corresponding indices. The number of clock cycles required for this stage is related to the number of quantized coefficients and the positions of uncoded states on the quantization paths.

Based on the intra-CG context template coding scheme, a four-channel parallel high-performance dependent quantization hardware architecture is designed in this paper, as illustrated in Fig. \ref{F.arch}. The quantization process is divided into eight stages, namely the pre-quantization stage (PQ), the distortion calculation and quantization candidate generation stages (M1–M3), the rate estimation stage (R1), the Viterbi pruning stage (V1), the context template update stage (U1), and the context index calculation stage (U2). Each stage requires one clock cycle to complete. In the figure, modules with the suffix “ff” represent registers. Gray and colored modules denote sequential logic circuits, while uncolored modules represent combinational logic circuits.

In the architecture, the PQ–M3 stages perform pre-quantization, distortion calculation, and quantization candidate generation. In the R1 stage, four Rate Est modules are instantiated to estimate the rates of the four quantization candidates and compute their rate–distortion costs by combining the estimated rates with the corresponding distortion values. The V1 stage implements the Viterbi algorithm, where paths with relatively larger accumulated rate–distortion costs are pruned. The U1 stage performs the context template update operation. According to the intra-CG context template scheme, only the quantization results within a single coefficient group (CG) are used for updating. As shown in the figure, four Update Template modules are instantiated to update the contextual information for the four quantization paths, respectively. In the U2 stage, the context indices are calculated based on the contextual information updated in the U1 stage. The resulting context indices are then fed into the rate estimation module.

To enable real-time quantization of ultra-high-definition videos, this hardware architecture features four key design points as follows:

(1) To improve the hardware clock frequency, the architecture splits the large-bit-width multiplications required for distortion calculation and quantization candidate value calculation into three clock cycles to complete (M1–M3). Other processes, including Pre-Quantization (PQ), Rate Estimation (R1), Viterbi Algorithm (V1), Context Updating (U1), and CtxInc Calculation (U2), are each finished in one clock cycle. Therefore, the process of quantizing a single transform coefficient takes 8 clock cycles in total.

(2) The rate estimation stage (R1) adopts the two schemes proposed in Chapter III: confining context templates within a CG and skipping $decAbslevel$ coding. At the cost of a slight BDBR loss, these schemes reduce memory access operations, allowing the rate estimation stage to be completed in one clock cycle.

\begin{figure*}[!t]
	\centering
	\includegraphics[width=7in]{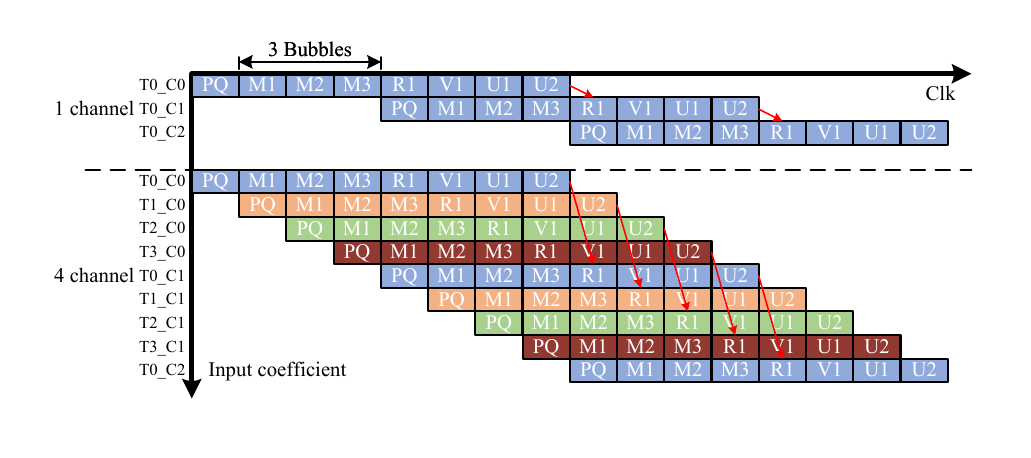}
	\caption{Timing diagram of single channel and 4-channel}
	\label{F.single_channel_and_4_channel}
\end{figure*}

(3) To improve throughput, the hardware architecture divides the single-coefficient quantization process into two phases: the first phase is PQ–M3, and the second phase is R1–U2, with each phase taking 4 clock cycles. Since the PQ–M3 phase has no dependency on context templates, the next coefficient can enter the PQ–M3 phase for computation while the previous coefficient is still in the R1–U2 phase. The timing diagram of this process is shown in the 1 channel section of Fig. \ref{F.single_channel_and_4_channel}, which means the throughput is improved to quantizing one transform coefficient every 4 clock cycles.

(4)To further enhance parallelism, this paper fully utilizes the waiting cycles during transform coefficient quantization on the basis of point (3), and proposes a 4-channel parallel hardware architecture for dependent quantization. The PQ–M3 process is designed with a pipelined structure, which enables a throughput of processing one transform coefficient per clock cycle in the PQ–M3 phase when quantizing 4 independent TBs. For the results output by the M3 phase per clock cycle, the R1–U2 phase introduces 4 sets of registers to reuse the same combinational logic for simultaneous quantization of transform coefficients from 4 independent TBs, as illustrated in the timing diagram of the 4 channel section of Fig. \ref{F.single_channel_and_4_channel}. As can be seen from the timing diagram in Fig. \ref{F.single_channel_and_4_channel}, during the 3 waiting cycles of the R1 phase of coefficient T0\_C1 until the completion of the U2 phase of coefficient T0\_C0, the quantization processes of T1\_C0, T2\_C0, and T3\_C0 are inserted. These coefficients are in different phases of the Quant Stage and do not interfere with each other.

The above four advantages enable the dependent quantization hardware architecture designed in this paper to achieve a throughput of quantizing one transform coefficient per clock cycle in the Quant Stage during the parallel quantization of multiple TBs. In addition, when the four TBs are independent of each other, the Output Stage can output four quantization results per clock cycle.

\section{Experimental Result and Analysis}

To analyze the performance metrics of the dependent quantization hardware architecture designed in this work, the proposed optimization scheme was implemented on VTM 18.0 to evaluate the BD-BR performance. In addition, the multi-channel parallel dependent quantization hardware architecture implemented in Verilog HDL was synthesized using Vivado and Design Compiler, respectively, to evaluate the hardware performance.

For the BDBR test experiments, the BDBR of the improved schemes proposed in this paper was measured under the All Intra and Random Access configurations, respectively. The 18 standard CTC sequences provided by JVET were adopted as the test videos, with the QP values set to 22, 27, 32 and 37. Under the All Intra configuration, the TemporalSubsampleRatio was set to 8 in accordance with the standard test requirements, meaning that one frame was encoded every 8 frames. To save test time under the Random Access configuration, 3 GOPs were tested for each video sequence, with each GOP containing 32 frames.

The BDBR experimental results are presented in Table \ref{T.BDBR_result}. As shown in Table \ref{T.BDBR_result}, compared with the standard DQ scheme, the proposed optimization scheme achieves BDBR increases of 0.42\% and 0.38\% under the All Intra and Random Access configurations, respectively. These results indicate that, at the same visual quality, the dependent quantization hardware scheme proposed in this work incurs only about a 0.4\% bitrate increase compared with the standard scheme, which is relatively marginal.

\begin{table}[!t]
	\caption{BDBR(\%) OF THE PROPOSED METHODS AGAINST VTM18.0 (All Intra And Random Access).}
	\label{T.BDBR_result}
	\centering
	\setlength{\tabcolsep}{4mm}{
		\begin{tabular}{c|c|c|c}
			\hline
			Class & Sequence & All Intra & Random Access \\
			\hline
			\multirow{3}{*}{A1} & Tango & 0.23 & 0.33 \\
			& FoodMarket & 0.33 & 0.17 \\
			& Campfire & 0.30 & 0.59 \\
			\hline
			\multirow{3}{*}{A2} & CatRobot & 0.42 & 0.28 \\
			& DaylightRoad & 0.41 & 0.41 \\
			& ParkRunning & 0.84 & 0.65 \\
			\hline
			\multirow{3}{*}{B} & RitualDance & 0.45 & 0.24 \\
			& Cactus & 0.39 & 0.44 \\
			& BasketballDrive & 0.45 & 0.34 \\
			\hline
			\multirow{4}{*}{C} & RaceHorsesC & 0.50 & 0.48 \\
			& PartyScene & 0.37 & 0.41 \\
			& BasketballDrill & 0.14 & 0.28 \\
			\hline
			\multirow{4}{*}{D} & RaceHorses & 0.71 & 0.43 \\
			& BQSquare & 0.39 & 0.28 \\
			& BlowingBubbles & 0.33 & 0.39 \\
			\hline
			\multirow{3}{*}{E} & FourPeople & 0.36 & - \\
			& Johnny & 0.54 & - \\
			& KristenAndSara & 0.38 & -\\
			\hline
			\multicolumn{2}{c|}{Average} & 0.42 & 0.38 \\
			\hline
	\end{tabular}}
\end{table}

To evaluate the maximum operating frequency and hardware resource consumption, the proposed high-parallel dependent quantization hardware architecture was implemented in Verilog HDL. Synthesis was then carried out using the Xilinx XCZU19 FPGA and the TSMC 28 nm process standard cell library. The synthesis results are presented in Table \ref{T.FPGA_syn_result} and Table \ref{T.ASIC_syn_result}.

As shown in Table \ref{T.FPGA_syn_result}, the proposed parallel dependent quantization hardware architecture consumes 55,575 LUTs, 55,974 FFs, 43 DSPs, and 8 BRAMs when implemented on the Xilinx XCZU19 FPGA, and the achievable operating frequency reaches 220 MHz.

As shown in Table \ref{T.ASIC_syn_result}, when synthesized using the TSMC 28 nm process standard cell library for ASIC implementation, the proposed parallel dependent quantization hardware architecture occupies an area of 0.69mm$^2$ with a power consumption of 370.125 mW, and the maximum operating frequency reaches 556 MHz.

\begin{table}[!t]
	\caption{FPGA Synthesis Results.}
	\label{T.FPGA_syn_result}
	\centering
	\setlength{\tabcolsep}{3mm}{
		\begin{tabular}{cc}
			\hline
			FPGA & XCZU19  \\
			\hline
			LUTs & 55575 \\
			FFs & 55974 \\
			DSPs & 43 \\
			BRAMs & 8 \\
			Frequency & 220MHz \\
			\hline
	\end{tabular}}
\end{table}

\begin{table}[!t]
	\caption{ASIC Synthesis Results.}
	\label{T.ASIC_syn_result}
	\centering
	\setlength{\tabcolsep}{3mm}{
		\begin{tabular}{cc}
			\hline
			ASIC & TSMC 28nm  \\
			\hline
			Area & 0.69mm$^2$ \\
			Power & 370.125mW \\
			Frequency & 556MHz \\
			\hline
	\end{tabular}}
\end{table}

To evaluate the quantization speed achievable by the proposed parallel dependent quantization hardware architecture, quantization data of 1000 32×32 transform blocks (TBs) were exported from six 4K video sequences using VTM 18.0. The exported quantization data were then fed into the proposed parallel dependent quantization hardware architecture implemented in Verilog HDL and simulated on the ModelSim platform. The number of clock cycles required for quantizing a 
32×32 TB was first measured. Based on this result, the number of cycles required for quantizing a single frame was further calculated. Finally, combined with the maximum operating frequency achievable by the hardware architecture, the frame per second (fps) for quantizing 4K video was obtained.

The clock cycles required by the proposed hardware architecture to quantize a 32×32 TB in the single-channel configuration under QP = 22, 27, 32, and 37 are listed in Table \ref{T.Quant32x32TB}. Considering that a 4K frame contains (3840×2160)/(32×32) TBs, the clock cycles required to quantize a 4K frame can be obtained by multiplying the results in Table \ref{T.Quant32x32TB} by (3840×2160)/(32×32). The calculated results are shown in Table \ref{T.QuantAFrame}.

\begin{table}[!t]
	\caption{Number of Cycles Required for Single-Channel Quantization of a 32×32 TB (clk).}
	\label{T.Quant32x32TB}
	\centering
	\setlength{\tabcolsep}{4mm}{
		\begin{tabular}{c|cccc}
			\hline
			Sequence & QP=22 & QP=27 & QP=32 & QP=37 \\
			\hline
			Tango & 3828 & 1182 & 303 & 208\\
			FoodMarket & 1918 & 364 & 89 & 71\\
			Campfire & 4275 & 3532 & 2369 & 1321\\ 
			CatRobot & 4392 & 2287 & 556 & 332\\
			DaylightRoad & 4839 & 3750 & 1139 & 640\\
			ParkRunning & 2598 & 2125 & 1675 & 1230\\
			\hline
	\end{tabular}}
\end{table}

\begin{table}[!t]
	\caption{Cycles Required for Single-Channel Quantization of One 4K Frame (Mclk).}
	\label{T.QuantAFrame}
	\centering
	\setlength{\tabcolsep}{4mm}{
		\begin{tabular}{c|cccc}
			\hline
			Sequence & QP=22 & QP=27 & QP=32 & QP=37 \\
			\hline
			Tango & 31.1 & 9.58 & 2.46 & 1.69\\
			FoodMarket & 15.53 & 2.95 & 0.72 & 0.57\\
			Campfire & 34.63 & 28.61 & 19.19 & 10.70\\ 
			CatRobot & 35.58 & 18.52 & 4.50 & 2.69\\
			DaylightRoad & 39.20 & 30.38 & 9.22 & 5.19\\
			ParkRunning & 21.05 & 17.21 & 13.57 & 9.97\\
			\hline
	\end{tabular}}
\end{table}

Based on Table \ref{T.QuantAFrame} and Equation \ref{E.fpsCalc}, the single-channel quantization throughput, i.e., the fps, is calculated, and the results are summarized in Table \ref{T.QuantAFrameFps}.

\begin{equation}
	\begin{aligned}
		fps = \frac{maxFreq}{clkCycle}.
	\end{aligned}
	\label{E.fpsCalc}
\end{equation}

$maxFreq$ represents the maximum operating frequency that the hardware can achieve, while $clkCycle$ denotes the number of clock cycles required for the hardware architecture to quantize one frame of 4K video.

\begin{table}[!t]
	\caption{Single-Channel Quantization Throughput(fps).}
	\label{T.QuantAFrameFps}
	\centering
	\setlength{\tabcolsep}{4mm}{
		\begin{tabular}{c|cccc}
			\hline
			Sequence & QP=22 & QP=27 & QP=32 & QP=37 \\
			\hline
			Tango & 7.1 & 23.0 & 89.5 & 130.3\\
			FoodMarket & 14.2 & 74.5 & 306.9 & 384.1\\
			Campfire & 6.4 & 7.7 & 11.5 & 20.6\\ 
			CatRobot & 6.2 & 11.9 & 48.9 & 81.7\\
			DaylightRoad & 5.6 & 7.2 & 23.9 & 42.4\\
			ParkRunning & 10.5 & 12.8 & 16.2 & 22.1\\
			Average & 8.3 & 22.8 & 82.8 & 113.5\\
			\hline
	\end{tabular}}
\end{table}

Since the four channels in the proposed parallel dependent quantization hardware architecture are identical and operate independently, the quantization speed of the 2-channel and 4-channel configurations can be obtained by multiplying the single-channel speed by two and four, respectively. The 4-channel configuration represents the maximum quantization speed achievable by the proposed architecture. The results are shown in Table \ref{T.FPGA_QuantSpeed} and Table \ref{T.ASIC_QuantSpeed}.

\begin{table}[!t]
	\caption{FRAME PER SECOND (Xilinx XCZU19 FPGA).}
	\label{T.FPGA_QuantSpeed}
	\centering
	\setlength{\tabcolsep}{3mm}{
		\begin{tabular}{c|cccc}
			\hline
			Number of channels & QP=22 & QP=27 & QP=32 & QP=37 \\
			\hline
			2-channel & 16.6 & 45.7 & 165.6 & 227.1\\
			4-channel & 33.2 & 91.4 & 331.2 & 454.1\\ 
			\hline
	\end{tabular}}
\end{table}

\begin{table}[!t]
	\caption{FRAME PER SECOND (TSMC 28nm ASIC).}
	\label{T.ASIC_QuantSpeed}
	\centering
	\setlength{\tabcolsep}{3mm}{
		\begin{tabular}{c|cccc}
			\hline
			Number of channels & QP=22 & QP=27 & QP=32 & QP=37 \\
			\hline
			single-channel & 21.0 & 57.7 & 209.2 & 286.9\\
			2-channel & 42.0 & 115.5 & 418.5 & 573.8\\
			4-channel & 84.0 & 231.0 & 837.0 & 1147.7\\ 
			\hline
	\end{tabular}}
\end{table}

As shown in Table \ref{T.FPGA_QuantSpeed} and Table \ref{T.ASIC_QuantSpeed}, the proposed high-parallel dependent quantization hardware architecture achieves quantization speeds of 4K@33.2, 91.4, 331.2, and 454.1 fps at QP = 22, 27, 32, and 37, respectively, when implemented on the Xilinx XCZU19 FPGA. When implemented on ASIC using the TSMC 28 nm process technology, the corresponding quantization speeds reach 4K@84.0, 231.0, 837.0, and 1147.7 fps.

To demonstrate the performance advantages of the proposed parallel dependent quantization hardware architecture, the experimental results are compared with those reported in two other studies on dependent quantization hardware architectures under the same FPGA synthesis conditions. The comparison results are shown in Table \ref{T.compareWithOthers}.

\begin{table}[!t]
	\caption{Comparison with Hardware Quantization Schemes.}
	\label{T.compareWithOthers}
	\centering
	\setlength{\tabcolsep}{1.5mm}{
		\begin{tabular}{c|ccc}
			\hline
			Design & Sheng et al.\cite{R:Q. SHENG} & Zhang et al.\cite{R:J. Zhang} & Proposed  \\
			\hline
			FPGA & XCZU19 & XCZU19 & XCZU19 \\
			Reference software & VTM14.2 & VTM18.0 & VTM-18.0\\
			BDBR(AI) & - & 0.81\% & 0.42\% \\ 
			BDBR(RA) & 1.05\% & 0.85\% & 0.38\% \\
			LUTs & 60586 & 8843 & 55575 \\
			FFs & 15301 & 10623 & 55974 \\
			DSPs & 42 & 43 & 43 \\
			BRAMs(36K) & 0 & 4 & 8 \\
			Frequency & 55.6MHz & 276MHz & 220MHz \\
			Speed(QP=22) & \multirow{4}{*}{1080P@60fps} & 4K@31.4fps & 4K@33.2fps \\
			Speed(QP=27) & & 4K@83.8fps & 4K@91.4fps \\
			Speed(QP=32) & & 4K@164.5fps & 4K@331.2fps \\
			Speed(QP=37) & & 4K@242.8fps & 4K@454.1fps \\
			\hline
	\end{tabular}}
\end{table}

From Table \ref{T.compareWithOthers}, it can be observed that the BDBR performance of the proposed scheme is superior to that of the schemes proposed by Sheng et al. \cite{R:Q. SHENG} and Zhang et al. \cite{R:J. Zhang} Specifically, compared with the scheme of Zhang et al. \cite{R:J. Zhang}, the BDBR loss is reduced by 0.39\% (0.81\% - 0.42\%) and 0.47\% (0.85\% - 0.38\%) under the All Intra and Random Access configurations, respectively. Moreover, under the Random Access configuration, the BDBR loss is reduced by 0.67\% (1.05\% - 0.38\%) compared with the scheme of Sheng et al. \cite{R:Q. SHENG} This improvement mainly benefits from the intra-CG context template scheme proposed in this paper, which is able to utilize more accurate contextual information during the rate estimation process.

In terms of hardware resource consumption, it can be observed that the proposed scheme increases the FF usage by 3.7× and 5.3× compared with the schemes proposed by Sheng et al. \cite{R:Q. SHENG} and Zhang et al. \cite{R:J. Zhang}, respectively. This is because the proposed architecture supports up to four-channel parallel quantization, which requires a large number of FF resources to simultaneously store the rate tables and contextual information for the four channels. Regarding LUT resources, the proposed scheme consumes slightly fewer LUTs than the scheme proposed by Sheng et al. \cite{R:Q. SHENG}, but significantly more than that of Zhang et al. \cite{R:J. Zhang}. This is mainly because the proposed scheme introduces a more complex context index calculation method based on the scheme of Zhang et al. \cite{R:J. Zhang} Although this design improves the BDBR performance, it comes at the cost of increased hardware resource consumption.

\begin{figure*}[!t]
	\centering
	\includegraphics[width=1\textwidth]{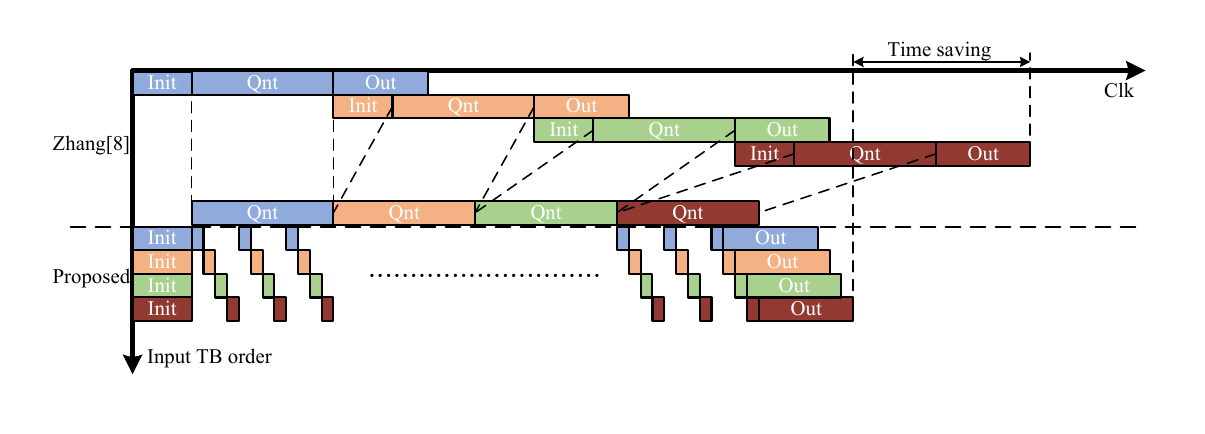}
	\caption{Comparison of Quantization Timing Diagrams}
	\label{F.HardSpeedComp}
\end{figure*}

\begin{figure*}[htb]
	\centering
	\includegraphics[width=0.6\textwidth]{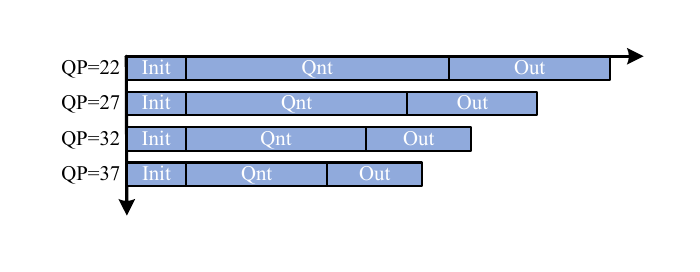}
	\caption{Effect of QP Reduction on Quantization Timing}
	\label{F.QPEffect}
\end{figure*}

In terms of quantization speed, it can be observed that the proposed scheme achieves higher quantization speeds than the scheme proposed by Zhang et al. at all QP values, and the improvement becomes more significant as the QP value increases. At QP = 37, the proposed scheme achieves nearly twice the quantization speed for 4K video compared with the scheme of Zhang et al. The improvement in quantization speed mainly results from the fact that the time required for the quantization stage in both architectures is identical. However, in the proposed scheme, when multiple TBs are quantized simultaneously, the initialization stage and the optimal quantization path output stage can be executed fully in parallel, as illustrated in Fig. \ref{F.HardSpeedComp}.

As can be seen from Fig. \ref{F.HardSpeedComp}, for the same four TBs to be quantized, the total latency incurred in the quantization stage (Qnt) is identical for our proposed method and that of Zhang et al., while the distinction lies in that parallelism is enabled in both the initialization (Init) and the final output (Out) stages within our design. At low-QP settings, where a greater number of coefficients are submitted to the Qnt stage, the share of cycles consumed by this stage among the total cycles per TB is raised. Owing to the ping‑pong buffer arrangement adopted by Zhang et al., the Out stage of one TB is, in most cases, superimposed with the Qnt stage of other TBs, thereby consuming almost no quantizer time. Consequently, only a slight improvement in quantization speed is attained by our proposal under low-QP settings.

The quantization speed of the proposed scheme increases with larger QP values because a higher QP results in a larger quantization step size, which causes the high-frequency zeroing process to set more coefficients to zero. Consequently, fewer coefficients enter the quantization stage for processing. As a result, when quantizing a TB, the initialization stage accounts for a significantly larger proportion of the total quantization time, as illustrated in Fig. \ref{F.QPEffect}. Under this condition, the dependent quantization hardware architecture proposed in this paper achieves a much higher quantization speed than the hardware architecture proposed by Zhang et al.\cite{R:J. Zhang}.

Based on the above analysis, the proposed DQ hardware parallel architecture achieves a significant improvement in BDBR performance and quantization speed at the cost of a fivefold increase in hardware resource consumption.

\section{Conclusion}
In this paper, to address the issues of poor parallelizability and inferior real-time performance in Dependent Quantization for H.266/VVC, we propose a highly parallel DQ hardware architecture with low BDBR performance loss, which adopts the intra-CG context and the decAbslevel skipping scheme proposed in this paper in its design. To verify the feasibility of the proposed scheme, we implemented it on the VTM18.0 software and tested its BDBR performance. Experimental results show that the proposed scheme incurs a mere BDBR loss of 0.42\% and 0.38\% under the All Intra and Random Access configurations, respectively. In addition, we implemented the proposed DQ hardware architecture using Verilog HDL, tested its quantization speed, and performed logic synthesis on the architecture with the FPGA and ASIC. The experimental results show that the proposed high-parallel dependent quantization hardware architecture achieves significantly better BDBR performance than the other schemes at the cost of substantially increased hardware resource consumption. Moreover, the quantization speeds can reach 4K@33.2, 91.4, 331.2, and 454.1 fps at QP = 22, 27, 32, and 37, respectively, when implemented on the Xilinx XCZU19 FPGA. When implemented on ASIC using the TSMC 28 nm process technology, the corresponding speeds reach 4K@84.0, 231.0, 837.0, and 1147.7 fps.

\newpage

\vspace{2cm}
\begin{IEEEbiography}[{\includegraphics[width=1in,height=1.25in,clip,keepaspectratio]{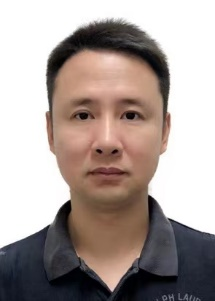}}]{Jun Zhang}
	received the Ph.D. degree in machine intelligence and systems engineering from Tohoku University, Japan, in 2004. During 2004-2011, he has worked for Fujitsu Limited, where he has engaged in the research of 3D graphics processor for automotive applications. Since 2011, he is a professor in Central South University, China. His research interests include the hardware architecture of chips for video coding, graphics processing units, AI acceleration, and robot control.
\end{IEEEbiography}
\vspace{1cm}
\begin{IEEEbiography}[{\includegraphics[width=1in,height=1.25in,clip,keepaspectratio]{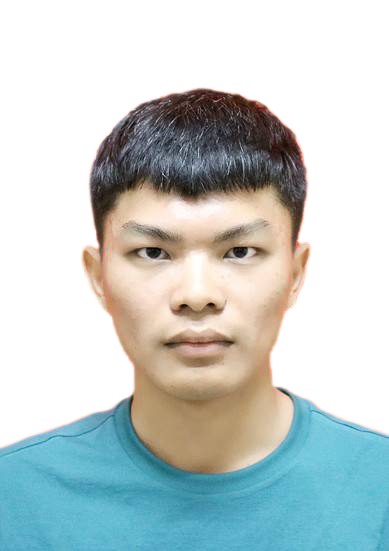}}]{Weizhi Bian}
	received the B.S. degree in electronic and information engineering from the Central South University, Changsha, China in 2023. He is currently pursuing the M.S. degree at the School of Automation, Central South University, Changsha, China. His research interests include the hardware architecture of chips for video coding.
\end{IEEEbiography}
\vspace{1cm}
\begin{IEEEbiography}[{\includegraphics[width=1in,height=1.25in,clip,keepaspectratio]{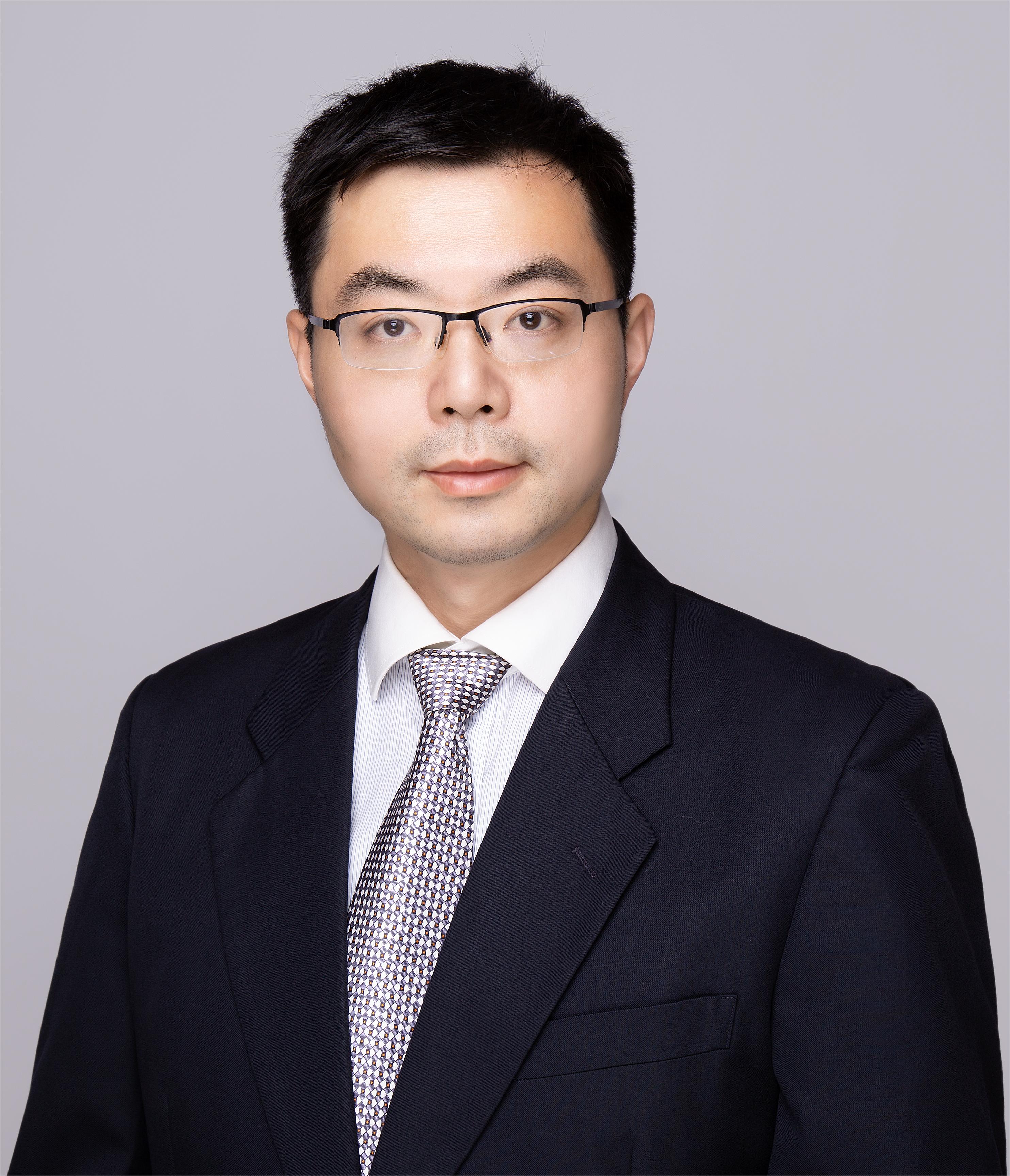}}]{Hao Zhang}
	received the Ph.D. degree in electrical engineering from Polytechnic University (now NYU Tandon School of Engineering), New York, NY, USA, in 2006. He is currently a professor with the School of Electronic Information, Central South University, Changsha, China. His recent research interests include video coding, image processing and sign language translation.
\end{IEEEbiography}

\vfill


\begin{thebibliography}{1}
\bibliographystyle{IEEEtran}

\bibitem{R:B. Bross}
B. Bross, Y.-K. Wang, Y. Ye, S. Liu, J. Chen, G. J. Sullivan, and J.-R. Ohm, ``Overview of the versatile video coding (vvc) standard and its applications,'' \textit{IEEE Transactions on Circuits and Systems for Video Technology}, vol. 31, no. 10, pp. 3736–3764, 2021.

\bibitem{R:Block partitioning}
Y.-W. Huang, J. An, H. Huang, X. Li, S.-T. Hsiang, K. Zhang, H. Gao,
J. Ma, and O. Chubach, ``Block partitioning structure in the vvc standard,'' \textit{IEEE Transactions on Circuits and Systems for Video Technology},
vol. 31, no. 10, pp. 3818–3833, 2021.

\bibitem{R:Intra prediction}
J. Pfaff, A. Filippov, S. Liu, X. Zhao, J. Chen, S. De-Luxán-Hernández,
T. Wiegand, V. Rufitskiy, A. K. Ramasubramonian, and G. Van der
Auwera, ``Intra prediction and mode coding in vvc,'' \textit{IEEE Transactions on Circuits and Systems for Video Technology}, vol. 31, no. 10, pp. 3834-3847, 2021.

\bibitem{R:Inter prediction}
J. Zhang, Y. Zhang, and H. Zhang, ``An efficient hardware architecture
of integer motion estimation based on early termination and data reuse
for versatile video coding,'' \textit{Expert Systems with Applications}, vol. 242, p. 122706, 2024.

\bibitem{R:Transform}
X. Zhao, S.-H. Kim, Y. Zhao, H. E. Egilmez, M. Koo, S. Liu, J. Lainema,
and M. Karczewicz, ``Transform coding in the vvc standard,'' \textit{IEEE
	Transactions on Circuits and Systems for Video Technology}, vol. 31,
no. 10, pp. 3878–3890, 2021.

\bibitem{R:Quantization}
H. Schwarz, M. Coban, M. Karczewicz, T.-D. Chuang, F. Bossen,
A. Alshin, J. Lainema, C. R. Helmrich, and T. Wiegand, ``Quantization
and entropy coding in the versatile video coding (vvc) standard,'' \textit{IEEE
	Transactions on Circuits and Systems for Video Technology}, vol. 31,
no. 10, pp. 3891–3906, 2021.

\bibitem{R:Loop Filter}
M. Karczewicz, N. Hu, J. Taquet, C.-Y. Chen, K. Misra, K. Andersson,
P. Yin, T. Lu, E. Francois, and J. Chen, ``Vvc in-loop filters,'' \textit{IEEE
	Transactions on Circuits and Systems for Video Technology}, vol. 31,
no. 10, pp. 3907–3925, 2021.

\bibitem{R:H. Lee}
H. Lee, S. Yang, Y. Park, and B. Jeon, ``Fast quantization method with
simplified rate–distortion optimized quantization for an hevc encoder,'' \textit{ IEEE Transactions on Circuits and Systems for Video Technology}, vol. 26, no. 1, pp. 107–116, 2016.

\bibitem{R:T. Nguyen Canh}
T. Nguyen Canh, M. Xu, and B. Jeon, ``Rate-distortion optimized quantization: A deep learning approach,'' 2018.

\bibitem{R:J. Cui}
J. Cui, S. Wang, S. Wang, X. Zhang, S. Ma, and W. Gao, ``Hybrid
laplace distribution-based low complexity rate-distortion optimized
quantization,'' \textit{IEEE Trans Image Process}, vol. 26, no. 8, pp. 3802-3816, 2017. [Online]. Available: https://www.ncbi.nlm.nih.gov/pubmed/28500003

\bibitem{R:M. Xu.RDOQ}
M. Xu, T. Nguyen Canh, and B. Jeon, ``Simplified level estimation for
rate-distortion optimized quantization of hevc,'' \textit{IEEE Transactions on
	Broadcasting}, vol. 66, no. 1, pp. 88–99, 2020.

\bibitem{R:M. Xu.DQ}
M. Xu and B. Jeon, ``Complexity-efficient dependent quantization for versatile video coding,'' \textit{IEEE Transactions on Broadcasting}, vol. 69, no. 3, pp. 832–839, 2023.

\bibitem{R:M. Wang}
M. Wang, S. Wang, J. Li, L. Zhang, Y. Wang, S. Ma, and S. Kwong,
``Low complexity trellis-coded quantization in versatile video coding,'' \textit{
	IEEE Transactions on Image Processing}, vol. 30, pp. 2378–2393, 2021.

\bibitem{R:Q. SHENG}
Q. SHENG, Y. CHENG, X. HUANG, C. LAI, X. HUANG, and H. YIN,
``A vvc dependent quantization optimization based on the parallel
viterbi algorithm and its fpga implementation,'' \textit{IEICE Transactions on
	Information}, vol. E107-D, pp. 797–806, 2024.

\bibitem{R:J. Zhang}
J. Zhang, W. Bian and H. Zhang, "A Low-Latency, Highly-Pipelined Hardware Architecture for H.266/VVC Dependent Quantization," in IEEE Transactions on Circuits and Systems I: Regular Papers, vol. 72, no. 8, pp. 4040-4051, Aug. 2025.

\bibitem{R:M. W.}
M. W. Marcellin and T. R. Fischer, ``Trellis coded quantization of
memoryless and gauss-markov sources,'' \textit{IEEE transactions on communications}, vol. 38, no. 1, pp. 82–93, 1990.

\bibitem{R:H. Schwarz}
H. Schwarz, T. Nguyen, D. Marpe, and T. Wiegand, ``Hybrid video
coding with trellis-coded quantization,'' in \textit{2019 Data Compression
	Conference (DCC)}. IEEE, Conference Proceedings, pp. 182–191.

\end{thebibliography}
\end{document}